# *Essential Metadata for 3D BRAIN Microscopy*

Alexander J. Ropelewski[1], Megan A. Rizzo[2], Jason R. Swedlow[3], Jan Huisken[4], Pavel Osten[5], Neda Khanjani[6], Kurt Weiss[4], Vesselina Bakalov[7], Michelle Engle[7], Lauren Gridley[7], Michelle Krzyzanowski[7], Tom Madden[7], Deborah Maiese[7], Justin Waterfield[7], David Williams[7], Carol Hamilton[7], and Wayne Huggins[7]

[1] Biomedical Applications Group, Pittsburgh Supercomputing Center, 300 S Craig Street, Pittsburgh, PA, 15213, USA
[2] Department of Physiology, University of Maryland School of Medicine, 660 West Redwood Street, Baltimore, MD, 21201, USA
[3] Centre for Gene Regulation & Expression, Division of Computational Biology, University of Dundee, Nethergate, Dundee, Scotland, DD1 4HN, United Kingdom
[4] Morgridge Institute for Research, 330 N Orchard Street, Madison, WI, 53715, USA
[5] Cold Spring Harbor Laboratory, One Bungtown Road, Cold Spring Harbor, NY, 11724, USA
[6] Mark and Mary Stevens Neuroimaging and Informatics Institute, Laboratory of Neuro Imaging, Keck School of Medicine of University of Southern California, 1975 Zonal Avenue, Los Angeles, CA, 90033, USA
[7] Bioinformatics and Computational Biology Program, RTI International, 3040 East Cornwallis Road, Research Triangle Park, NC, 27709, USA

## ABSTRACT
Recent advances in fluorescence microscopy techniques and tissue clearing, labeling, and staining provide unprecedented opportunities to investigate brain structure and function. These experiments' images make it possible to catalog brain cell types and define their location, morphology, and connectivity in a native context, leading to a better understanding of normal development and disease etiology. Consistent annotation of metadata is needed to provide the context necessary to understand, reuse, and integrate these data. This report describes an effort to establish metadata standards for 3D microscopy datasets for use by the Brain Research through Advancing Innovative Neurotechnologies® (BRAIN) Initiative and the neuroscience research community. These standards were built on existing efforts and developed with input from the brain microscopy community to promote adoption. The resulting Essential Metadata for 3D BRAIN Microscopy includes 91 fields organized into seven categories: Contributors, Funders, Publication, Instrument, Dataset, Specimen, and Image. Adoption of these metadata standards will ensure that investigators receive credit for their work, promote data reuse, facilitate downstream analysis of shared data, and encourage collaboration.

## INTRODUCTION
New fluorescence microscopy techniques, coupled with recent advances in tissue clearing, labeling, and staining provide unprecedented opportunities to investigate brain structure and function. These technologies have been used to generate high-resolution, three-dimensional (3D) images of entire brains from model organisms[1-8] and large sections of brains from humans[9,10]. Data from these images have been used for cell-type-specific mapping of the brain connectome[11], identifying and characterizing brain regions governing behavior[12-14], and defining structural aspects of disease pathologies[9]. The number of 3D microscopy datasets will soon grow exponentially with the introduction of major brain initiatives from countries around the world. The images from these efforts will make it possible to catalog brain cell types and define their location, morphology, and connectivity in a native context, leading to a better understanding of normal development and disease etiology.

To maximize the utility of these data, it will be essential to establish and adopt standards for annotating, reporting, and formatting imaging datasets. Such standards can facilitate scientific transparency, rigor, and reproducibility; aid data sharing and integration; and promote the development of analysis tools and a sustainable informatics ecosystem. There are several standards efforts relevant for microscopy that are currently underway. Example community standards include minimum information guidelines for fluorescence microscopy developed within the 4D Nucleome Project[15], minimal metadata and data structures developed within the National Institutes of Health (NIH) Common Fund's Stimulating Peripheral Activity to Relieve Conditions (SPARC) project (https://sparc.science/help/3FXikFXC8shPRd8xZqhjVT), data structures for 2D and 3D microscopy being developed under the Brain Imaging Data Structure Extension Proposal 031, and metadata guidelines from Recommended Metadata for Biological Images[16]. Other relevant efforts include Quality Assessment and Reproducibility and Images in Light Microscopy, which aims to improve the overall quality and reproducibility of microscopy datasets[17], the Global BioImaging initiative, which is establishing recommendations and guidelines for image data repositories and formats[18], and the International Neuroinformatics Coordinating Facility (INCF), which coordinates neuroinformatics infrastructure and standards[19].

This article describes an effort to establish metadata standards for 3D microscopy datasets for use by the Brain Research through Advancing Innovative Neurotechnologies® (BRAIN) Initiative and larger neuroscience research community. The metadata standards were developed by a 10 member Working Group (WG; https://doryworkspace.org/WorkingGroupRoster) with diverse expertise using a consensus-based process that relied on input from the scientific community. The metadata standards build on existing efforts including the DataCite metadata schema[20] and the Open Microscopy Environment (OME)[21] to promote adoption and utility. The resulting Essential Metadata for 3D BRAIN Microscopy is designed to ensure that a 3D microscopy dataset is sufficiently described to support its reuse by scientists who did not generate the data. These metadata standards are being implemented within the Brain Image Library (BIL)[22], the designated repository to accept and make microscopy data publicly available for investigators funded by the BRAIN Initiative. Adoption of these metadata standards will aid investigators who want to share data, helping them to evaluate and decide which data can be combined. The use of these metadata standards also ensures that datasets comply with FAIR (**F**indable, **A**ccessible, **I**nteroperable, and **R**eusable) principles[23] and the policy for data sharing for the BRAIN Initiative[24].

## RESULTS

The Essential Metadata for 3D BRAIN Microscopy includes 91 metadata fields organized into seven categories: Contributors, Funders, Publication, Instrument, Dataset, Specimen, and Image. Each metadata field is specified by a name, a definition, a list of allowable values, whether it is required or optional for submission to the BIL (https://www.brainimagelibrary.org), and the number of times it can be repeated for a dataset. The complete metadata specification is available from the Defining Our Research Methodology (DORy) website (https://doryworkspace.org/metadata).

To encourage adoption and reduce burden on data submitters, only 31 of the fields in Essential Metadata for 3D BRAIN Microscopy are required for submission of a dataset to the BIL.

Nineteen of the required metadata fields support assignment of a Digital Object Identifier (DOI) to the dataset. A DOI is a unique code that provides a permanent and stable mechanism for retrieval of a dataset and its metadata[25]. The remaining required fields are necessary for investigators to open and view the images and to determine if they want to reuse the dataset. A summary of the metadata fields in each category is presented in Table 1.

**Table 1. Essential Metadata for 3D BRAIN Microscopy Required Fields**

| Field Name | Definition | Allowable Values | Support DOI? |
|---|---|---|---|
| **Contributors Category (nine required metadata fields)** | | | |
| contributorName | Person (last name, first name) or organization (e.g., research group, department, institution) contributing to or responsible for the project, but does not include funders of the project. If a contributor has more than one contributorType, use a separate line for each. | Free text | Yes |
| Creator | Main researchers involved in producing the data. There must be at least one creator. | Yes<br>No | Yes |
| contributorType | Categorization of the role of the contributor. Recommended: ProjectLeader (for principal investigator), ResearchGroup (for lab, department, or division). | ContactPerson; DataCollector DataCurator; ProjectLeader; ProjectManager; ProjectMember; RelatedPerson; Researcher; ResearchGroup; Other | Yes |
| nameType | Type of contributorName. | Organizational<br>Personal | Yes |
| nameIdentifier | Alphanumeric code that uniquely identifies an individual or legal entity, (listed in the contributorName field). Accepted identifiers include GRID, ISNI, ORCID, ROR, and RRID. Preferred identifiers are ORCID for personal names and ROR for organizational names. Required for Personal nameType. | Free Text | Yes |
| nameIdentifier Scheme | Identifying scheme used in nameIdentifier. Required for Personal nameType. | GRID*; ISNI*; ORCID*; ROR*; RRID* | Yes |
| affiliation | Organizational or institutional affiliation of the contributor. | Free text | Yes |
| affiliationIdentifier | Unique identifier (ROR preferred) for the organizational or institutional affiliation of the contributor. | Free text | Yes |
| affiliationIdentifier Scheme | Identifying scheme used in affiliationIdentifier. | GRID*; ISNI*; ORCID*; ROR*; RRID* | Yes |
| **Dataset Category (five required metadata fields)** | | | |
| Title | Short phrase by which the specific dataset is known (e.g., title of a book). | Free text | Yes |
| Rights | Any rights information for the dataset. May be the name of the license and can include embargo or other use restrictions on data (see https://spdx.org/licenses). | Free text | Yes |
| rightsURI | If using a common license and licensing information is online, provide a link to the license. | Free text | Yes |
| rightsIdentifier | If using a common license, provide the Software Package Data Exchange (SPDX) code for the license (see https://spdx.org/licenses). | Free text | Yes |
| Abstract | Additional descriptive information about the dataset, including a brief description and the context in which it was created (e.g., aim of the experiment, what the dataset is expected to show). This abstract will be used on the Digital | Free text | Yes |

|  | Object Identifier (DOI) landing page and will be the primary description of the dataset; it will ideally be 100+ words. | | |
|---|---|---|---|
| **Funders Category (five required metadata fields)** | | | |
| funderName | The name of the funder. | Free text | Yes |
| fundingReferenceIdentifier | Alphanumeric code that uniquely identifies an individual or legal entity. Preferred identifier is ROR. | Free text (or URL) | Yes |
| fundingReferenceIdentifierType | Identifying scheme used in fundingReferenceIdentifier. | GRID*; ISNI*; ORCID*; ROR*; RRID* | Yes |
| awardNumber | Funding code or project number assigned to the grant. | Free text | Yes |
| awardTitle | Title of the grant award. | Free text | Yes |
| **Instrument Category (two required metadata fields)** | | | |
| MicroscopeType | Type of microscope used to capture the image (e.g., inverted, upright, light sheet, confocal, two photon). | Free text | No |
| MicroscopeManufacturerAndModel | Manufacturer and model of the microscope used. | Free text | No |
| **Image Category (five required metadata fields)** | | | |
| xAxis | Predominant tissue direction as one moves from the left side of the image to the right side of the image. | Left to right; Right to left; Anterior to posterior; Posterior to anterior; Inferior to superior; Superior to inferior; Oblique | No |
| yAxis | Predominant tissue direction as one moves from the top of the image to the bottom of the image. | Left to right; Right to left; Anterior to posterior; Posterior to anterior; Inferior to superior; Superior to inferior; Oblique | No |
| zAxis | Predominant tissue direction as one follows a given pixel position through the stack of images from the first image to the last image. | Left to right; Right to left; Anterior to posterior; Posterior to anterior; Inferior to superior; Superior to inferior; Oblique | No |
| Number | Number assigned to each channel. | Free text | No |
| displayColor | Display color of each channel in triplet (red, green, blue) format. | Free text | No |
| **Specimen Category (five required metadata fields)** | | | |
| Species | Common organism classification name for the donor organism (e.g., mouse, human). | Free text | No |
| NCBITaxonomy | National Center for Biotechnology Information (NCBI) taxonomy code for species of the donor organism. | Free text | No |
| Age | Age of the donor (or unknown). | Free text | No |
| Ageunit | Unit for the age of the donor. | Days; Months; Years | No |
| Sex | Sex of the donor. | Male; Female; Unknown | No |

*For more information on the identifiers, see the Global Research Identifier Database (GRID; https://www.grid.ac), International Standard Name Identifier (ISNI; https://isni.org), Open Researcher and Contributor ID (ORCID; https://orcid.org), Research Organization Registry (ROR; https://ror.org), and Research Resource Identifiers (RRID; https://scicrunch.org/resources) websites.

*Contributors.* The Contributors category includes nine required metadata fields that identify and give credit to the scientists and organizations involved in the creation of the dataset (https://doryworkspace.org/metadata, Table 1). Contributors include the broader set of researchers and institutions (except funders) that participated in the development of the dataset (e.g., data collection, management, and/or distribution). Each contributor is identified by name, role on the project, and affiliation. The Creator field is used to indicate whether a contributor is also a creator ("Yes" or "No"). Creators are the principle researchers involved in the generation of the dataset. There must be a least one creator for each dataset. All nine metadata fields in the Contributors category, including controlled vocabulary and references to digital identifiers (e.g., ORCID, RRID), are equivalent to properties in the DataCite metadata schema and support assignment of a DOI to the dataset[20].

*Dataset.* The Dataset category includes 15 metadata fields that provide a high-level description of the data including title, abstract, methods, imaging modality, and how the data can be reused (https://doryworkspace.org/metadata). The WG defines a dataset as a collection of images (in one or more files) generated from a light microscope that can be stacked to form a distinct 3D volume or object, such as an entire brain or a brain region. Relevant 3D light microscope technologies include (but are not limited to) scanning confocal microscopy (point or line scanning), spinning disk confocal microscopy, multiphoton excitation microscopy (2p or 3p), light sheet microscopy (selective/single plane, oblique, ultramicroscopy), structured illumination, and deconvolved widefield. Five of the Dataset metadata fields are required, including title, abstract, and those describing the intellectual property rights (Table 1). Optional items include fields to describe the general modality (e.g., morphology, connectivity), technique (e.g., anterograde tracing, smFISH), and methods used to generate biological materials and computationally process the data. Eight of the metadata fields in the Dataset category are equivalent to properties in the DataCite metadata schema and support assignment of a DOI to the dataset[20].

*Funders.* The Funders category includes five metadata fields that describe the organizations providing financial support for the generation of the dataset (https://doryworkspace.org/metadata, Table 1). The five Funders metadata fields are only required if the project is funded by government agencies, such as the NIH. The WG recognized that some information (e.g., award number) may not be available for projects funded by foundations and internal mechanisms. All the Funders metadata fields, including uniquely identifying organizations (e.g., ORCID, RRID), are equivalent to properties in the DataCite metadata schema and support assignment of a DOI to the dataset[20].

*Instrument.* The Instrument category includes 12 metadata fields that describe the instrument used to capture the images (https://doryworkspace.org/metadata). There are two required fields to record the microscope type and model (Table 1). Optional fields record the details of the objective, detector, type of illumination and wavelength, and temperature of the sample. All fields in the instrument category take free text, but there are suggested values for describing the objective immersion medium, type of illumination, and type of detector. All the Instrument metadata fields are equivalent to properties in the OME Data Model[21].

*Image.* The Image category includes 33 metadata fields that describe the size and 3D orientation of the image, channels and fluorophores used, and location of relevant landmarks (https://doryworkspace.org/metadata). Seven of the Image metadata fields are required, including those describing the x, y, and z orientation, channel numbers, colors, and number of microns per pixel in the x and y dimensions (Table 1). Optional fields include items to describe oblique dimensions (if applicable), the name and coordinates of landmarks, the number of pixels in the x, y, and z dimension, and the number of files, timepoints, channels, and slices. Fifteen of the metadata fields in the Image category are equivalent to properties in the OME Data Model[21]. Exceptions include those fields describing oblique dimensions, landmarks, and the number of channels and slices.

*Specimen.* The Specimen category includes 12 metadata fields that describe the donor, organ, and sample being studied (https://doryworkspace.org/metadata). Five of the Specimen metadata fields are required, including those to record the common name, National Center for Biotechnology Information taxonomy code, age, and sex of the organism being studied (Table 1). Optional fields include items to capture the genotype of the organism, organ name, location or region where the sample is found, and name of the atlas used to describe the location (if applicable).

*Publication.* The Publication category includes five optional metadata fields that identify publications, preprints, and protocols that are related to the dataset (https://doryworkspace.org/metadata). Each publication, preprint or protocol can be identified by a globally unique identifier, a PubMed Central identifier if applicable, and/or a citation. Three of the Publication metadata fields, including controlled vocabulary to identify the relationship of the publication, preprint, or protocol to the dataset, are equivalent to properties in the DataCite metadata schema and support assignment of a DOI to the dataset[20].

**Implementation of the Essential Metadata in the Brain Image Library**

The BIL is an NIH-funded national public resource enabling researchers to deposit, analyze, mine, share, and interact with large brain image datasets[22]. BIL is the designated repository to accept and make microscopy data publicly available for investigators funded by the BRAIN Initiative. The BIL's process for collecting metadata currently consists of having the depositor fill out a metadata spreadsheet that is then uploaded through a submission portal and attached to a dataset. The existence of the BIL predates this standard, thus the existing metadata collected within the BIL is only a small subset of the metadata described here.

The BIL is implementing and will require the use of these metadata standards by all depositors in an upcoming portal release. To prepare for this implementation, the BIL piloted the collection of metadata aligned with the standard among a subset of existing data contributors. Within this pilot, investigators were asked to review and expand a multitabbed metadata spreadsheet that was prepopulated with information from their submitted datasets along with the existing BIL metadata mapped onto the new metadata schema. From the pilot, it was found that the data contributors could provide the minimally requested information, but there were areas within the collection spreadsheet where the instructions were unclear and misinterpreted. These instructions have been clarified and are reflected in this publication. In addition, feedback was received concerning reducing repetitive entries in the spreadsheet (e.g., the Contributors and Funders categories which are frequently the same for all datasets submitted by a project). DOIs for the datasets in the pilot will be available in late spring 2021.

## Discussion

To promote adoption and maximize opportunities for data integration, the Essential Metadata for 3D BRAIN Microscopy builds on existing metadata schemas from DataCite[20] and the OME Data Model[21,26]. DataCite (https://datacite.org) is a nonprofit organization that provides DOIs for a wide variety of research outputs, including datasets. To assign a DOI to a dataset, DataCite requires a specified list of metadata fields and a digital location (e.g., URL) where the dataset and metadata can be accessed. Twenty-seven of the 90 metadata fields described here, including 19 of the 31 required fields (Table 1), are equivalent to properties in the DataCite metadata schema. These properties are either required or recommended for DataCite to assign a DOI. Once created, DataCite makes the DOI and metadata publicly available, and datasets can be indexed and searched using common platforms like Google. Together, these characteristics ensure that microscopy datasets are **F**indable, **A**ccessible, and **R**eusable according to FAIR principles[27]. Additionally, publications that provide a direct link to a dataset in a repository (e.g., by a DOI) are correlated with higher citation impact[28].

OME (https://docs.openmicroscopy.org/ome-model/latest) is a well-established informatics framework for storing and sharing biological microscopy data that is widely used in the US and international research communities. Forty-one of the 90 metadata fields within the Essential Metadata for 3D BRAIN Microscopy are either equivalent or can be mapped to terms in the OME metadata schema. The OME Data Model[21] is being used in the Bioimaging in North America Network (https://www.bioimagingna.org) and several related large-scale microscopy projects, including the SPARC project and the 4D Nucleome project. Alignment of the Essential Metadata for 3D BRAIN Microscopy with OME will help ensure that datasets in the BIL are interoperable with other high-priority microscopy projects and efforts.

This approach, combining metadata fields from DataCite and OME, balances attribution and retrieval of datasets with the specificity necessary for annotating microscopy datasets. Additional fields were added to the Essential Metadata for 3D BRAIN Microscopy to account for data management issues and to address specific use cases in brain microscopy.

Next steps will focus on building tools and features to support metadata submission to the BIL and extending the utility and promoting the adoption of the Essential Metadata for 3D BRAIN Microscopy. Within the BIL, plans include reducing the data entry burden and improving consistency of collected metadata. This includes collecting Contributors and Funders metadata within the BIL submission portal itself and creating a mechanism to attach this metadata to a series of datasets. This improvement would help groups, such as the BRAIN Initiative Cell Census Network (BICCN), that have investigators depositing new entries over a period of months or years. There is also the possibility to reduce data entry further by building lookup capabilities into the portal that might, for example, query external portals that have restful APIs (such as the NIH Research Portfolio Online Reporting Tool Expenditures and Results [RePORTER][29]) or project-specific resources on GitHub.

To promote data interoperability and integration, mapping the Essential Metadata for 3D BRAIN Microscopy to additional relevant standards and associating metadata fields and response options with standard vocabularies and ontologies will be explored. The metadata standards will also be submitted to the INCF for review. INCF endorsement will promote dissemination to the international scientific community and help ensure that future development is coordinated with other standards.

Future efforts will also include developing additional metadata fields to describe experimental techniques relevant for 3D brain microscopy. The Essential Metadata for 3D BRAIN Microscopy establishes a baseline set of concise, low-burden metadata fields that are flexible and broadly applicable enough to represent the variety of techniques present among imaging datasets in the BIL. Additional metadata for experimental techniques will help investigators reproduce experiments, evaluate image quality, identify artefacts, and determine whether datasets are suitable for integration and comparison.

## Methods

The Essential Metadata for 3D BRAIN Microscopy was developed by the BRAIN Initiative 3D Microscopy WG between October 2019 and August 2020.
The WG consisted of 10 members from the BICCN and the larger scientific community. WG members' expertise included informatics, 3D microscopy, neuroscience, physiology, engineering and spanned data generation, data archiving, data integration, and data analysis.

This expertise helped ensure that the WG represented the perspective and needs of the BRAIN Initiative and the larger neuroscience research community.

The WG held an initial in-person meeting in October 2019 to define the scope of the standard and to establish the framework for prioritizing metadata attributes. During this meeting, the WG discussed the needs and perspectives of the user community and heard presentations on the BRAIN Initiative, BICCN, BIL, and OME. The WG discussed several key factors to be considered in the development of the Essential Metadata for 3D BRAIN Microscopy. First, the standards should support the goals of the BRAIN Initiative and the larger neuroscience research community. Second, the standards should build on existing standards efforts. Third, a subset of metadata should be required for all imaging datasets. Finally, the standards should be flexible enough to address the diversity of microscopy techniques within the BICCN and larger research community, including those not yet invented.

Following the in-person meeting, a metadata subgroup was established to develop an initial set of metadata fields and attributes. Between November 2019 and January 2020, the subgroup reviewed experimental modalities and technologies and specified use cases for querying, understanding, and reusing microscopy datasets. The subgroup also reviewed related metadata standardization efforts, including the OME[21], Neuroimaging Data Model[30], Brain Imaging Data Structure[31], Force11[32], and DataCite[20]. Relevant metadata fields were captured in a spreadsheet that included the name of the metadata field, category, subcategory (if applicable), occurrence (whether the metadata field is required and how many times it can appear for each dataset), whether the metadata field is needed to generate a DOI, definition, units (if applicable), and allowable values (e.g., free text or controlled vocabulary lists).

The metadata subgroup presented its initial recommendations to the WG during an in-person meeting in February 2020. The WG provided feedback on the proposed metadata fields and attributes, including which fields should be required and whether any should be added, removed, or clarified. The WG also recommended that future efforts include additional metadata fields to address reproducibility and quality control. At the end of the meeting, the WG approved the preliminary draft of the standard.

The preliminary metadata standards were posted on the DORy website for feedback and comment by the larger research community between April 21 and May 12, 2020. Potential reviewers were identified by searching the NIH RePORTER[29] and the BRAIN Initiative website[33] for funded projects that included keywords for relevant microscopy techniques. These investigators were sent an email describing the project and asking them to provide feedback. Respondents were asked to review the metadata fields and attributes and complete Likert-style scales to indicate whether the proposed standards were sufficient and would be easy to adopt and implement as well as to provide general comments and suggestions.

The WG reviewed outreach results during a teleconference in May 2020. A total of 42 individuals provided comments on the preliminary standard. In general, the response to outreach was positive, and numerous helpful comments and suggestions were offered. Between June and August 2020, the metadata subgroup met several times to review community input and make updates before finalizing version 1.0 of the Minimal Archival Metadata Standard for 3D BRAIN Microscopy.

**Acknowledgements**
Research reported in this publication was supported by the National Institute of Mental Health (NIMH) of NIH under award number R24MH114683. The Brain Image Library is supported by the National Institutes of Mental Health of NIH under award number R24MH114793. The authors wish to acknowledge the members of the BRAIN Initiative 3D Microscopy Working Group: Alex Ropelewski (chair), Jan Huisken (chair), Hong-Wei Dong, Lydia Ng, Megan Rizzo, Jason Swedlow, Carol Thompson, Pavel Osten, Neda Khanjani and Kurt Weiss. The authors also thank Thien Lam for expert editorial assistance. The content is solely the responsibility of the authors and does not necessarily represent the official views of NIH, NIMH, or any of the sponsoring organizations and agencies, or the U.S. government.

**Author Contributions**
W.H. was the primary author on the manuscript. A.R. led the development and refinement of the standards and contributed the description of their implementation within the BIL. A.R. N.K., M.A.R., J.S., J.H., P.O., and K.W. contributed to the manuscript either by their participation in the Working Group and/or by editing or commenting on the manuscript text. V.B., M.E., L.G., M.K., T.M., D.M., J.W., D.W., and C.H. contributed to the manuscript by supporting the Working Group consensus process and/or by editing and commenting on the manuscript.

**Competing Interests**
JRS is Founder and CEO of Glencoe Software, Inc. which builds imaging data solutions for academic, biotech and pharma R&D labs.